# Growth diagram of La$_{0.7}$Sr$_{0.3}$MnO$_3$ thin films using pulsed laser deposition

Hangwen Guo,[1,2] Dali Sun,[2,3] Wenbin Wang,[1,2] Zheng Gai,[2,4] Ivan Kravchenko,[4] Jian Shao,[5] Lu Jiang,[1,2] Thomas Z. Ward,[2] Paul C. Snijders,[2] Lifeng Yin,[5] Jian Shen*,[1,5] and Xiaoshan Xu*[2]

[1]*Department of Physics and Astronomy, University of Tennessee, Knoxville, TN 37996, USA*

[2]*Materials Science and Technology Division, Oak Ridge National Laboratory, Oak Ridge, TN 37831, USA*

[3]*Department of Physics and Astronomy, University of Utah, Salt Lake City, UT 84112*

[4]*Center for Nanophase Materials Sciences, Oak Ridge National Laboratory, Oak Ridge, TN 37831, USA*

[5]*State Key Laboratory of Surface Physics and Department of Physics, Fudan University, Shanghai 200433, China*

**Abstract**: An experimental study was conducted on controlling the growth mode of La$_{0.7}$Sr$_{0.3}$MnO$_3$ thin films on SrTiO$_3$ substrates using pulsed laser deposition (PLD) by tuning growth temperature, pressure and laser fluence. Different thin film morphology, crystallinity and stoichiometry have been observed depending on growth parameters. To understand the microscopic origin, the adatom *nucleation*, *step advance* processes and their relationship to film growth were theoretically analyzed and a growth diagram was constructed. Three boundaries between highly and poorly crystallized growth, 2D and 3D growth, stoichiometric and non-stoichiometric growth were identified in the growth diagram. A good fit of our experimental observation with the growth diagram was found. This case study demonstrates that a more comprehensive understanding of the growth mode in PLD is possible.

## I. INTRODUCTION

The pulsed laser deposition (PLD) has become one of the most popular techniques in epitaxial film growth because of its versatility and conceptual simplicity.[1–6] However, the physical processes of PLD are far from simple, because it involves multiple steps including the ablation of target material, the plasma generation and propagation in background gas, the deposition of ablated atoms on substrates, and the non-equilibrium processes on the surface such as diffusion, desorption, nucleation and attaching onto existing atomic steps.[7–10] Clearly, it is of great importance to understand how these microscopic processes and the corresponding growth parameters determine the way the films grow (i.e. growth mode), which has significant effects on the physical properties of the films.

Previously, the individual growth parameters, such as substrate temperature, background-gas pressure, and laser fluence have been shown to affect the growth properties greatly. For example, Infortuna Harvey and Gauckler reported a systematic study of yttrium-stabilized zirconia (YSZ) and cerium gadolinium oxide (CGO) thin films grown on various substrates by PLD. It was found that high background-gas ($O_2$) pressure favors porous morphology as opposed to the dense structure.[11] Kan and Shimakawa studied the effect of laser fluence on the growth of $BaTiO_3$ thin film on $SrTiO_3$ substrates. The results show that low laser fluences make the stoichiometry of the films deviate from that of the target and in turn affect the ferroelectric properties.[12] The crystalline properties of $SrTiO_{3-\delta}$ homoepitaxial thin films have been reported by Ohtomo and Hwang.[13] The growth phase diagram in terms of $O_2$ pressure and temperature has been studied. It is noted that at the high temperature and low pressure region, the mismatching between crystallization and oxidation timescale gives rise to irregular *nucleation* and growth cycle. Furthermore, the temperature is found to have a critical influence on the diffusion barrier. The diffusivity increase exponentially with increasing temperature, which may strongly affect the growth mode.[1,9,14]

In addition, great interests have been raised on the behavior of the growth-mode boundaries, which have not been fully understood. For example, the boundary between the 2D and the 3D growth mode is still under debate. Metev et al. experimentally defined this boundary using substrate temperature and deposition rate as parameters[1,15], while different dependence of the boundary on growth parameters is proposed theoretically by Hong et al. comparing the lifetime of adatoms diffusion and landing time interval.[16] Besides, the relation between growth parameters and the degree of crystallinity and stoichiometry still remains unclear. A growth diagram which can describe different growth modes under different growth conditions[13,14,16] will be desirable.

In this work, we resolve the aforementioned debate through a comprehensive experimental study and a theoretical treatment. We choose a prototypical system of La$_{0.7}$Sr$_{0.3}$MnO$_3$ (LSMO) thin films grown on SrTiO$_3$ (STO) (001) substrates to study the growth properties under different conditions. This composition of LSMO is a known half-metallic ferromagnet above room temperature which exhibits interesting properties such as spin-polarized tunneling and colossal magnetoresistance.[17] Such features offer great potential in many applications, such as spin valves and resistive random access memories (RRAM).[17–23] We thermodynamically analyzed the microscopic processes of adatom *nucleation, step advance* and their influence on the growth properties of oxide thin films, to answer the questions of how to decide the boundary between growth modes. Based on our model analysis, we propose a growth diagram which describes the dependence of surface morphology, crystallinity and stoichiometry of oxide films on supersaturation and substrate temperature as the growth parameters. Our experimental observations of the LSMO growth modes under various growth conditions fit in this growth diagram nicely. We emphasize supersaturation as an important concept in constructing and understanding growth diagrams for PLD thin film growth of complex oxides.

This paper is organized as follows. Section II describes the experimental conditions used in this work. Section III presents experimental results including the dependence of thin film growth mode on temperature, pressure and laser fluence. Section IV discusses the theoretically constructed growth diagram and its comparison with our experimental data.

## II. EXPERIMENTAL METHODS

Thin films of $La_{0.7}Sr_{0.3}MnO_3$ were grown on single crystal substrates of $SrTiO_3$ (001) (in-plane lattice mismatch of 0.8%[24]) using PLD with a KrF ($\lambda$= 248 nm) laser in an oxygen background containing 10% ozone,[25,26] with the growth pressure of 2 mTorr. The typical film thickness is 10 nm. The repetition rate of the laser is kept at 1 Hz. *In-situ* high-pressure Reflection High Energy Electron Diffraction (RHEED) is used to monitor the entire deposition process. The time dependence of the intensity of the specular reflection was recorded. Here we employ the Pulse Laser Interval Deposition technique (see Ref. [2]), i.e. the growth periods are separated by the annealing periods (typically 5 min) in which laser pulses are paused. Specifically, the laser is paused every time when the RHEED intensity reaches a local maximum to allow sample annealing which manifests itself as the upturn of RHEED intensity. The laser pulse is resumed when the RHEED intensity saturates. The laser fluence was varied between 1 and 4 $J/cm^2$. The substrate temperature was varied from 650 to 840 $^o$C. The substrates are treated by buffered-HF and pre-annealed in $O_2$ (1 atm) for 3 hours at 950 $^o$C. The target-substrate distance was 4 cm. The RHEED images were taken using KSA-400 camera at exposure time of 667 ms at the end of the growth. *Ex-situ* Atomic Force Microscopy (AFM) is used to obtain details of surface morphology. X-ray diffraction (XRD) experiments are conducted to measure the crystallinity of the films. Energy-dispersive X-ray spectroscopy (EDX) is carried out to analyze the chemical composition.

### III. EXPERIMENTAL RESULTS

A. Temperature dependence

We first study the influence of growth temperature while fixing the background oxygen pressure and the laser fluence at 1J/cm$^2$. Figure 1a shows AFM and RHEED images of the pretreated SrTiO$_3$ substrates. Flat surfaces with terrace width of approximately 300 nm are obtained. The sharp RHEED pattern also indicates the presence of flat terraces and a good surface crystallinity. Figure 1b and 1c show the *ex-situ* room temperature AFM, RHEED oscillations and *in-situ* RHEED images at growth temperature of 720 $^o$C and 776 $^o$C, respectively. In Fig. 1b, for the sample grown at 720 $^o$C, the single layer terraces inherited from the substrate are still visible while many sub-monolayer-height islands exist on top of each terrace. The RHEED intensity can only maintain its initial level for the first several oscillations, after which it decreases upon further growth. The final RHEED pattern shows weak intensity contrast, although no extra diffraction spots are present. The room temperature AFM data indicates that the growth mode is still 2D layer by layer at 720 $^o$C, with a RMS roughness of only 0.14 nm. The measured X-ray Diffraction rocking curve shows the full width of half maximum (FWHM) of 0.046$^o$, compared to 0.039$^o$ for a well-crystallized film.[27] These features suggest that the film surface is poorly crystallized with 2D layer-by-layer feature preserved. In Fig. 1c, the room temperature AFM image of the sample grown at 776 $^{o}$C reveals neither regular terraces nor steps from the substrate and has rms roughness of 0.45 nm. The RHEED intensity decreases during the growth and displays a weak pattern with extra spots, indicating 3D island formation. For sample grown at 840$^{o}$C, as shown in Fig. 1d, totally different features are present. Rod-like structures (as high as 100 nm) are clearly visible in the AFM image. EDX experiments were conducted to investigate the stoichiometry of these features. Fig. 1d shows the Mn-Kα and O-K spectroscopy maps. The Mn and O concentrations of the film and the rod are different, indicating chemical phase separation occurs in this growth regime resulting in different stoichiometry of the rods as compared to the film. As shown in Table 1, a semiquantitative analysis[28] indicates that the Mn:O composition ratio is roughly 1:1 on the rod and 1:3 on the film. The Sr concentration is hard to determine due to large background from the STO substrate. We cannot resolve the La concentration due to its proximity to the substrate Ti peak. According to the data in TABLE I, the composition on the rod is likely to be MnO due to the high vapor pressure of Sr and chemical bond stability of MnO.[29]

The temperature dependence of growth mode has also been investigated at a different laser fluence of 4J/cm$^2$. As shown in Fig. 2a, sample grown at 660 $^{o}$C have 2D layer-by-layer features (AFM and RHEED images will be discussed in Section II-C). The sample grown at 760 $^{o}$C shows a completely different morphology. In Fig. 2b-d, a 3D growth mode is revealed in both AFM and RHEED images. At first, the maximum intensity of primary RHEED spot gets weaker with each oscillation, then it remains nearly unchanged with almost no oscillations. The RHEED pattern shows spots with a strong intensity, indicating a well crystallized, 3D growth mode.

B. Pressure dependence

Next, we tune the background oxygen pressure while keeping the temperature (730 $^{o}$C) and laser fluence (1 J/cm$^2$) constant. A tube was installed with its end close to the sample surface to supply an oxygen gas pressure during the growth process. To ensure the same laser and temperature conditions, we use a larger substrate (5mm×8mm) and carefully position the oxygen tube nozzle with respect to the sample to create a pressure gradient over the surface of one sample. Similar approach has been used to create temperature gradient by other authors.[30] As shown in Fig. 3a, a 3D growth mode is obtained at low oxygen pressure, while flatter films are visible at higher pressure. In Fig. 3b, we show the rms roughness of the entire film area as measured by *ex-situ* AFM. It is clearly seen that the rms roughness decreases with the direction of increasing local pressure. To verify that the roughness distribution does not originate from temperature variations due to the large substrate size, we moved our substrate by 1 mm laterally with respect to the oxygen tube and grew another sample under the same conditions. We find that the rms roughness distribution on the sample correspondingly shifts by about 1 mm, which excludes temperature non-uniformity effects as the cause for the observed surface roughness distribution.

C. Laser fluence dependence

In addition to the temperature and the oxygen pressure, the pulsed laser fluence is another important parameter to control the film quality. Here we keep the temperature at 740 $^\circ$C and increase the laser fluence from 1 J/cm$^2$ to 2.2 J/cm$^2$. Dramatic changes are observed in the film growth as shown in Fig. 4a and 4b. The sample grown at 1 J/cm$^2$ shows 3D features with FWHM of 0.050$^\circ$ in the rocking curve[27], which indicates the poorly-crystallized nature. The sample grown at 2.2 J/cm$^2$ indicates a 2D layer-by-layer growth mode with FWHM of 0.039$^\circ$. According to Fig. 4c and 4d, perfect oscillations as well as good RHEED patterns are observed, both indicating good surface morphology and crystallinity.

III.    Discussion

As a starting point, the change in surface energy $\Delta\sigma = \sigma + \sigma_i - \sigma_s$ is widely adopted to analyze observed surface morphology, where $\sigma$, $\sigma_s$ and $\sigma_i$ denote the surface energy of the depositing layer, the underlying layer and the interface energy, respectively.[10] Indeed, such analysis lies in the core of the origin of the well-known crystal growth mechanisms such as Volmer-Weber (VW)[31], Frank-van der Merwe (FM)[32] and Stranski-Krastanov (SK)[33] growth. When $\Delta\sigma < 0$, 2D layer-by-layer growth mode (FM) is favored; while 3D islands growth mode (VW) is preferred when $\Delta\sigma > 0$. The SK growth mode represents a transition from 2D to 3D growth when the lattice strain is taken into account. However, this description of growth is oversimplified. For example, even for $\Delta\sigma > 0$, it is still possible to achieve 2D layer-by-layer growth.[10] As seen above, by tuning experimental parameters, different varieties of surface morphology, surface crystallinity and stoichiometry have been observed. As shown in Fig.5a, the microscopic growth process involves multiple steps, more than the picture of surface energy change. Moreover, what also needs to be described is the relation between the growth conditions and the crystallinity and the stoichiometry of the films. Note that the discussion of stoichiometry is even beyond the picture of Fig. 5a where the smallest components are the unit cells.

In this paper, we applied the theoretical treatment in Ref. [10] on the PLD growth of oxide thin films, i.e. 1) to consider the factors of particle exchange and energy barrier in nucleation to account for the crystallization; 2) to consider the rate of *step advance* not only in terms of the diffusion process, but also the adatom concentration and their spatial gradient on the surface which all play important roles in determining the boundary between 2D and 3D growth modes. In addition, the film stoichiometry is described using the supersaturation of the corresponding vapor to solid process.

Here we first focus on the discussion on the temperature dependence of the nucleation process which strongly affects the crystallinity of thin films. Then, we focus on the discussion on the competition between *nucleation* and *step advance* (the growth of 2D islands and advance of steps), which determines the boundary of 2D layer-by-layer and 3D growth. A growth phase diagram is developed based on those discussions in terms of growth temperature and supersaturation which is a useful concept in describing the growth conditions. Our experimental findings and theoretical model are compared to test the feasibility of our growth diagram.

A. Theoretical Growth Diagram

During PLD growth, the laser ablation generates a large atomic flux. As illustrated in Fig.5 (a), these incoming atoms become adatoms on the substrate surface and diffuse. Some coalesce and become nuclei. At low laser repetition rate,[34] the steady-state rate of *nucleation* can be described as[27]:

$$J_{nuc} = f(\Delta\mu, T)\exp\left(-\frac{H(\Delta\mu)}{kT}\right) \qquad (1)$$

$$f(\Delta\mu, T) = 2\sqrt{s_c}\, arN_0\left(\frac{\Delta\mu - s_c\Delta\sigma}{\pi kT}\right)^{1/2} \qquad (2)$$

$$H(\Delta\mu) = \frac{4\chi^2 s_c}{\Delta\mu - s_c\Delta\sigma} - (E_{des} - E_{sd}) \qquad (3)$$

where $s_c$ is the area of the surface unit cell; $a$ is the lattice constant; $N_0$ is the density of adsorption sites; $\chi$ is the step edge energy per unit length; $r$ is the arrival rate which is proportional to the concentration of adatoms; $k$ is Boltzmann constant; $E_{des}$ and $E_{sd}$ are the desorption and diffusion energy barriers. $\Delta\mu$, known as supersaturation, is the chemical potential difference of adatoms transitioning from their quasi-vapor phase (the mobile adatoms on the surface and the background oxygen in the gas phase) near the substrate to their solid phase on the substrate.

The factor $f(\Delta\mu, T)$ can be considered as an effective Zeldovich factor which accounts for the deviation of the system from the equilibrium state; it describes the rate of atom exchange between the nuclei and its quasi-vapor parent phase. The factor $H(\Delta\mu)$ denotes the energy barrier of the *nucleation*. The competition between $f(\Delta\mu, T)$ and $\exp(-H(\Delta\mu)/kT)$ as a function of temperature results in a maximum value of *nucleation* rate:

$$T_{nuc}^m = H(\Delta\mu)/2k \qquad (4)$$

When the sample temperature is low, the *nucleation* rate is low because it's difficult to overcome the *nucleation* energy barrier. At the same time, the effective

Zeldovich factor is relatively high, indicating a low atom exchange rate between gas and solid. Thus, the *nucleation* rate is low and the films are not well crystallized. When the temperature is high, the *nucleation* rate is limited due to a low Zeldovich factor. However, it is easier to overcome the *nucleation* energy barrier to form nuclei. Thus, the films are well crystallized although the *nucleation* rate is also low. Consequently, Eq. (4) divides the boundary between the poorly crystallized and well crystallized growth modes, as illustrated in Fig. 5(b).

As shown in Fig. 5(a), besides *nucleation*, another way for adatoms to contribute to the film growth is to attach to existing nuclei or steps causing *step advance*. The process includes the surface diffusion of adatoms towards the steps or edges of nuclei and incorporation of the adatoms into the kinks.[27] Assuming that the growth is in the diffusion region, i.e. the diffusion process is the limiting factor of step advance, the rate of step-advance can be written in the form of[10]

$$V_{sa} = 2a\nu \frac{\Delta \mu}{kT} \exp\left(-\frac{\varphi - (E_{des} - E_{sd})/2}{kT}\right) \quad (5)$$

where $\nu$ is the vibrational frequency of the adatom; $\varphi$ is the adsorption energy at the kink position. We note that similar to *nucleation*, the rate of step-advance is also a function of supersaturation and temperature.

The annealing process used in this work for each monolayer helps to reach 2D layer by layer growth in the later stage of a monolayer deposition. For example, the small nucleation on top of 2D islands may become unstable due to the lowered supersaturation without laser pulses and decompose into adatoms which eventually attach to the step edges of the lower layer via interlayer mass transfer. Thus, the processes at the early stage of a monolayer deposition such as *nucleation* and surface migration of the adatoms are more important to determine the growth properties.[2] In such cases, the competition between the *nucleation* rate and the *step advance* rate becomes a crucial factor to determine the growth mode. As illustrated in Fig. 5(c), if the *nucleation* rate $J_{nuc}$ is much higher than the rate of *step advance* $V_{sa}$, new nuclei can form on top of existing islands before the completion of the underlying layer. In turn, several layers can grow simultaneously, causing 3D growth. Such a growth mode induces a reduction of the peak intensity of RHEED oscillation. In the other case, if the *nucleation* rate is much lower than the rate of *step advance*, new nuclei will form after most of the underlying layer is filled, which gives rise to a 2D layer-by-layer growth mode.

To compare the timescales, we calculate:

$$t_{layer} = \frac{L}{V_{sa}}; \qquad (6)$$

$$t_{nuc} = \frac{1}{J_{nuc}L^2} \qquad (7)$$

where $t_{layer}$ denotes the time to completely cover the substrate terrace with width *L* (the upper limit of the distance of the step advance) by one monolayer via *step advance*; $t_{nuc}$ denotes the time to form one nucleus on the same substrate terrace.

The boundary in the growth diagram between 2D layer-by-layer or 3D island growth is given approximately by $t_{layer} \sim t_{nuc}$ :

$$\frac{L^3 \sqrt{S_C} r N_0}{\upsilon} \left( \frac{\Delta\mu - S_C \Delta\sigma}{\pi \Delta\mu^2} kT \right)^{1/2} \exp\left( \frac{\varphi + (E_{des} - E_{sd})/2 - \dfrac{4\chi^2 S_C}{\Delta\mu - S_C \Delta\sigma}}{kT} \right) \sim 1 \quad (8).$$

As can be seen in Eq. (8), 2D layer-by-layer growth can be achieved above a certain threshold $\Delta\mu$ even with $\Delta\sigma > 0$. In addition, the step width $L$ is also an important parameter to tune the growth modes.[16]

By considering all the above discussions, taking Eq. (4) and Eq. (8), using $\Delta\mu$ and $T$ as variables, we are able to construct a growth phase diagram, as shown in Fig. 6. For simplicity, a Kossel crystal (i.e. here a layer refers to a layer of unit cell instead of an atomic monolayer) has been considered here and we only consider the nearest neighbor interaction for the strength (or the bond energy) $b$. The parameters used are: $L$=300 nm; $\nu = 10^{13}$ Hz; a=0.4 nm; $r$=$10^{22}$ cm$^{-2}$s$^{-1}$; $E_{des} = 2b$; $E_{sd} = b$; $\chi = b/2$; and $\varphi = 3b$. The change of surface energy $\Delta\sigma$ is assumed to be zero.

The boundary $L_1$ (red online) corresponds to $\Delta\mu = 0$. Below $L_1$ the growth is non-stoichiometric due to the inability of completing the thermo-chemical transition from the quasi-vapor phase to solid phase of the certain compound.[35] Above $L_1$ the film can be grown with the right stoichiometry. Boundary $L_2$ is calculated using Eq. 8 which separates the 2D layer-by-layer growth from the 3D growth. Boundary $L_3$ is calculated using Eq. 4 which separates the poorly crystallized (P-C) and well crystallized (W-C) growth modes. Here we assume that all the boundaries are independent with each other. Thus, five different regions can be defined in the phase diagrams: non-stochiometric; poorly crystallized 3D (P-C 3D); poorly crystallized 2D layer-by-layer (P-C LBL); well crystallized 2D layer-by-layer (W-C LBL); well crystallized 3D (W-C 3D).

We note that the low laser repetition rate used (1 Hz) allows us to use steady-state considerations since the adatoms reach steady-state concentration during pulse intervals.[34] For simplification, we used the average deposition rate to analyze the process of *nucleation* and *step advance*. We also neglected the effect of epitaxial strain which affects step bunching phenomena in the step flow regime.[16]

B. Comparison of Experimental Results with the Growth Diagram

It is important to verify whether the predicted phase diagram is consistent with the experimental data, and provides useful guidance on thin film growth of complex oxides by PLD.

The supersaturation, though not a direct tunable experimental parameter, is dependent on the temperature $T$, background oxygen pressure $P$, and the laser fluence (i.e. atom arrival rate $r$).

For the dependence of $\Delta\mu$ on the oxygen background pressure, we consider the vapor-solid phase transition.[27] $\Delta\mu$ increases logarithmically with $P$:

$$\partial\Delta\mu(P,T)/\partial \ln P = RT \qquad (9).$$

Similarly, the dependence of $\Delta\mu$ on the arrival rate of material ablated from the target, which is proportional to the concentration of adatoms is:

$$\partial\Delta\mu(P,T)/\partial \ln r = RT \qquad (10).$$

When the oxygen pressure and arrival rate are constant, one can derive the temperature dependence of the $\Delta\mu$:[27]

$$\partial\Delta\mu(P,T)/\partial \ln T \approx -\Delta h \qquad (11)$$

where $\Delta h$ denotes the molar enthalpy change between the solid and vapor phase.

In Fig.7, we summarize our experimental AFM images and fit them into the theoretical growth diagram. For guiding purpose, we use different arrows to illustrate the qualitative dependence of growth mode under different growth parameters.

First, the temperature dependence of samples grown at a laser fluence of 1 J/cm$^2$ is shown (the dash-dot-dot arrow). As discussed, the sample grown at 720 $^o$C has a poorly crystallized 2D layer-by-layer feature, so it falls into P-C LBL region. By increasing the temperature to 776 $^o$C, the growth mode becomes 3D, corresponding to region P-C 3D; while further increasing the growth temperature to 840 $^o$C will lead the system to non-stoichiometry. Next, we study the temperature dependence of samples grown at relatively large laser fluence of 4 J/cm$^2$, as indicated by the dash-dot arrow. For sample grown at 660 $^o$C, the high supersaturation is able to put the system into a well crystallized 2D layer-by-layer growth (region W-C LBL). By increasing the temperature from 660 $^o$C to 760 $^o$C, the sample crosses into a well crystallized 3D growth mode (region W-C 3D), consistent with our theoretical understanding.

We also examine the effect of laser fluence at fixed temperature and background pressure. Higher laser fluence translates into a higher ablated atom arrival rate, which implies a larger supersaturation. Indeed our experiments reveal that a higher laser fluence can lead the samples from a poorly-crystallized 3D phase (region P-C 3D) into a well-crystallized 2D layer-by-layer phase (region W-C LBL), as indicated by long dash arrow in Fig.7. Similar results have been revealed in experiments in which films become smoother when increasing the laser repetition rate.[36,37]

The increase of the local oxygen pressure also corresponds to an enhancement of the supersaturation value according to Eq. 9, which changes the growth mode from P-C 3D to P-C LBL (solid arrow); such tendency corresponds to the decrease of rms roughness as observed in our experiment.

The experimental and theoretical results in this work confirms the findings of Metev et al.[1,15] in which a high deposition rate (high supersaturation) and a low growth temperature favored a 2D growth mode, while the unverified trend of the boundary between 2D step-flow growth and 3D island formation constructed by W. Hong et al.[16] theoretically appears to be inconsistent with our work. The observation of Ohtomo and Hwang[13] also fits nicely in our more complete growth diagram because according to our theoretical model, the supersaturation decreases with temperature and increases with oxygen pressure. The effect of background pressure and sample-target distance has been discussed by M. Koubaa et al.[38] where experimentally a wide range of surface morphology involving grains and column has been shown, though a typical 2D layer-by-layer growth mode is missing. While our experimental study on pressure effect focuses on a small range of pressure change near the optimal growth condition and helps on strengthening the comprehensive theoretical phase diagram.

**CONCLUSION**

To summarize, we studied the surface morphology, crystallinity and stoichiometry of LSMO thin films on STO (001) substrates grown using PLD. Various growth modes and phases have been observed. Theoretical considerations establish a growth phase diagram which reveals the nature of different growth modes in terms of supersaturation and temperature under the following condition: 1) the change of surface energy $\Delta\sigma$ is ignorable; 2) the *step advance* is in the diffusion region; 3) the early stage of forming a layer is the most important in the growth process. As a result of the thorough theoretical framework, our derived growth diagram excellently matches the experimentally observed growth modes. As a case study, our results demonstrate the possibility of more comprehensive understanding on controlling growth process and film qualities in PLD growth.

*Correspondence to: xiaoshan.xu@gatech.edu and shenj5494@fudan.edu.cn

**ACKNOWLEDGMENTS**

Research supported by the US Department of Energy, Basic Energy Sciences, Materials Sciences and Engineering Division (P.C.S., T.Z.W., X.S.X.) and performed in part at the Center for Nanophase Materials Sciences (CNMS) (Z.G., I.K.) User Facility, which are sponsored at Oak Ridge National Laboratory by the Office of Basic Energy Sciences, US Department of Energy. We also acknowledge partial funding supports from the National Basic Research Program of China (973 Program) under Grant No. 2011CB921801 (J.S.), and the US DOE Office of Basic Energy Sciences, the US DOE Grant No. DE-SC0002136 (H.W.G., W.B.W)

**REFERENCES:**

[1] D.B. Chrisey and G.K. Hubler, *Pulse Laser Deposition of Thin Films* (1994).

[2] H.M.C. and G. Eres, Journal of Physics: Condensed Matter **20**, 264005 (2008).

[3] A. Ohtomo and H.Y. Hwang, Nature **427**, 423 (2004).

[4] D.P. Norton, Materials Science and Engineering: R: Reports **43**, 139 (2004).

[5] H.W. Jang, D.A. Felker, C.W. Bark, Y. Wang, M.K. Niranjan, C.T. Nelson, Y. Zhang, D. Su, C.M. Folkman, S.H. Baek, S. Lee, K. Janicka, Y. Zhu, X.Q. Pan, D.D. Fong, E.Y. Tsymbal, M.S. Rzchowski, and C.B. Eom, Science **331**, 886 (2011).

[6] T.J. Jackson and S.B. Palmer, Journal of Physics D: Applied Physics **27**, 1581 (1994).

[7] P.R. Willmott and J.R. Huber, Rev. Mod. Phys. **72**, 315 (2000).

[8] P.R. Willmott, Progress in Surface Science **76**, 163 (2004).

[9] M.N.R. Ashfold, F. Claeyssens, G.M. Fuge, and S.J. Henley, Chemical Society Reviews **33**, 23 (2004).

[10] I. V Markov, *Crystal Growth for Beginners: Fundamentals of Nucleation, Crystal Growth and Epitaxy*, 2nd ed. (2003).

[11] A. Infortuna, A.S. Harvey, and L.J. Gauckler, Advanced Functional Materials **18**, 127 (2008).

[12] D. Kan and Y. Shimakawa, Applied Physics Letters **99**, 81903 (2011).

[13] A. Ohtomo and H.Y. Hwang, Journal of Applied Physics **102**, 83704 (2007).

[14] S.K. Sinha, R. Bhattacharya, S.K. Ray, and I. Manna, Materials Letters **65**, 146 (2011).

[15] S. Metev and K. Meteva, Applied Surface Science **43**, 402 (1989).

[16] W. Hong, H.N. Lee, M. Yoon, H.M. Christen, D.H. Lowndes, Z. Suo, and Z. Zhang, Physical Review Letters **95**, 95501 (2005).

[17] H.Y. Hwang, S.-W. Cheong, N.P. Ong, and B. Batlogg, Physical Review Letters **77**, 2041 (1996).

[18] J.D. Ferguson, Y. Kim, L.F. Kourkoutis, A. Vodnick, A.R. Woll, D.A. Muller, and J.D. Brock, Advanced Materials **23**, 1226 (2011).

[19] X. Hong, A. Posadas, A. Lin, and C.H. Ahn, Physical Review B **68**, 134415 (2003).

[20] S. Brivio, C. Magen, A.A. Sidorenko, D. Petti, M. Cantoni, M. Finazzi, F. Ciccacci, R. De Renzi, M. Varela, S. Picozzi, and R. Bertacco, Physical Review B **81**, 94410 (2010).

[21] D. Sun, L. Yin, C. Sun, H. Guo, Z. Gai, X.-G. Zhang, T.Z. Ward, Z. Cheng, and J. Shen, Physical Review Letters **104**, 236602 (2010).

[22] A. Chen, Z. Bi, C.-F. Tsai, L. Chen, Q. Su, X. Zhang, and H. Wang, Crystal Growth & Design **11**, 5405 (2011).

[23] A. Chen, Z. Bi, C.-F. Tsai, J. Lee, Q. Su, X. Zhang, Q. Jia, J.L. MacManus-Driscoll, and H. Wang, Advanced Functional Materials **21**, 2423 (2011).

[24] F. Tsui, M.C. Smoak, T.K. Nath, and C.B. Eom, Applied Physics Letters **76**, 2421 (2000).

[25] H.-Y. Zhai, J.X. Ma, D.T. Gillaspie, X.G. Zhang, T.Z. Ward, E.W. Plummer, and J. Shen, Physical Review Letters **97**, 167201 (2006).

[26] T.Z. Ward, S. Liang, K. Fuchigami, L.F. Yin, E. Dagotto, E.W. Plummer, and J. Shen, Physical Review Letters **100**, 247204 (2008).

[27]*See Supplementary Material as [URL Will Be Inserted by AIP] for More Detailed Information of Nucleation, Diffusion and Step Advance, Thermodynamics of Phase Transition and Features of Poorly & Well Crystallized Growth.*

[28]*X-Ray Data Booklet* (Lawrence Berkeley National Laboratory, 2009).

[29]http://www.veeco.com/pdfs/mbe/vapor-pressure-Chart-1.pdf .

[30] I. Ohkubo, H.M. Christen, S. V Kalinin, J. G. E. Jellison, C.M. Rouleau, and D.H. Lowndes, Applied Physics Letters **84**, 1350 (2004).

[31] M. Volmer and A. Weber, Zeitschrift Für Physikalische Chemie **119**, 277 (1926).

[32] F.C. Frank and J.H. van der Merwe, Proceedings of the Royal Society of London. Series A. Mathematical and Physical Sciences **198**, 205 (1949).

[33] I. Stranski and L. Krastanov, Sitzugnsber. Akad. Wissenschaft Wien **146**, 797 (1938).

[34] M. Kareev, S. Prosandeev, B. Gray, J. Liu, P. Ryan, A. Kareev, E.J. Moon, and J. Chakhalian, Journal of Applied Physics **109**, 114303 (2011).

[35] W. Wang, Z. Gai, M. Chi, J.D. Fowlkes, J. Yi, L. Zhu, X. Cheng, D.J. Keavney, P.C. Snijders, T.Z. Ward, J. Shen, and X. Xu, Phys. Rev. B **85**, 155411 (2012).

[36] G.H. Lee, B.C. Shin, and B.H. Min, Materials Science and Engineering: B **95**, 137 (2002).

[37] J. Shin, S. V Kalinin, A.Y. Borisevich, E.W. Plummer, and A.P. Baddorf, Applied Physics Letters **91**, 202901 (2007).

[38] M. Koubaa, A.M. Haghiri-Gosnet, R. Desfeux, P. Lecoeur, W. Prellier, and B. Mercey, Journal of Applied Physics **93**, 5227 (2003).

**Figure Captions**:

FIG. 1. (Color online) (a) AFM image (left) and RHEED image (right) of SrTiO$_3$ (001) substrate. (b), (c) AFM, real-time RHEED oscillations and RHEED images of samples grown at 720 $^o$C and 776 $^o$C under 1J/cm$^2$ laser fluence. (d) AFM (upper left) and SEM (upper right) images of sample grown at 840 $^o$C under 1J/cm$^2$ laser fluence. Mn (lower left) and O (lower right) EDX spectroscopy are shown corresponding to SEM image area.

FIG. 2. (Color online) Samples grown at different temperatures with laser fluence of 4J/cm$^2$. (a), (b) surface morphologies of samples grown at 660 $^o$C and 760 $^o$C. (c), (d) corresponding RHEED oscillations and image of sample in (b).

FIG. 3. (Color online) (a) growth pressure dependence of surface morphology under growth temperature of 730 $^o$C and laser fluence of 1J/cm$^2$ . (b) surface roughness (in unit of nm) map of the sample under oxygen pressure gradient. The dashed arrow indicates the direction of increasing local pressure.

FIG. 4. (Color online) Laser fluence dependence of samples grown at (a) 1J/cm$^2$ and (b) 2.2J/cm$^2$, both under growth temperature of 740 $^o$C. (c), (d) corresponding RHEED oscillations and image of the sample in (b).

FIG. 5. (Color online) Microscopic illustration of the growth processes. (a) Schematic

diagram of the atomic process in the deposition. (b) Schematic of the poor crystalization (left) and the good crystalization (right). (c) Schematic of the 2D layer-by-layer growth and the 3D growth.

FIG. 6. (Color online) The theoretically constructed growth diagram. $L_1$ (red): boundary between stoichiometric and non-stoichiometric growth; $L_2$ (green): boundary between 2D layer-by-layer (LBL) and 3D growth; $L_3$ (blue): boundary between the poorly crystallized (P-C) and well crystallized (W-C) growth.

FIG. 7. (Color online) Qualitative comparison between experimental results and theoretical growth diagram. Arrows description: dash-dot-dot: temperature increase (720 $^o$C, 776 $^o$C, 840 $^o$C) under constant laser fluence (1J/cm$^2$) and pressure; dash-dot: temperature increase (660 $^o$C, 760 $^o$C) under constant laser fluence (4J/cm$^2$) and pressure; long dash: laser fluence increase (1J/cm$^2$, 2.2 J/cm$^2$) under constant temperature and pressure; solid: pressure increase under constant temperature (730 $^o$C) and laser fluence (1J/cm$^2$).

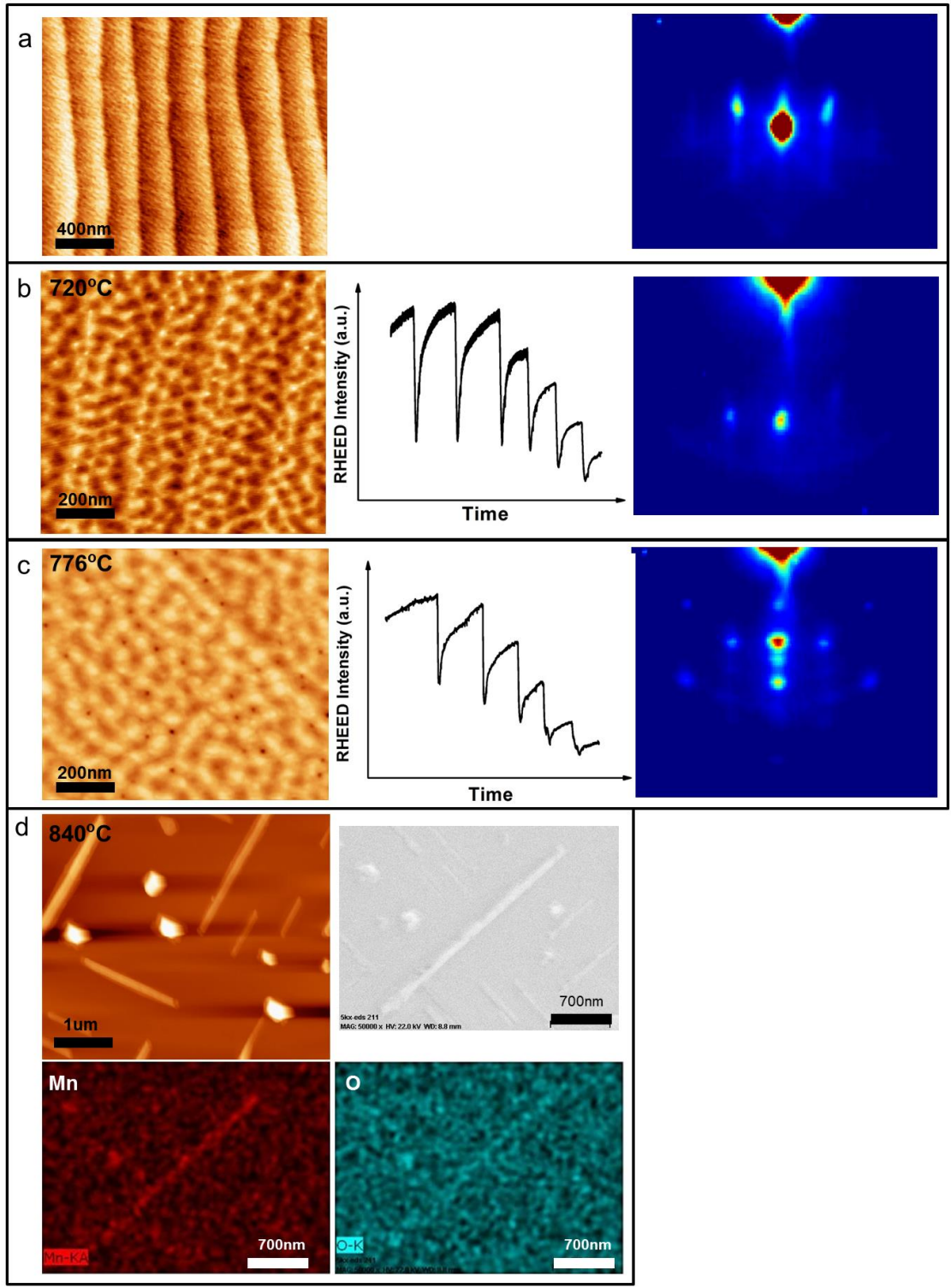

Figure 1.

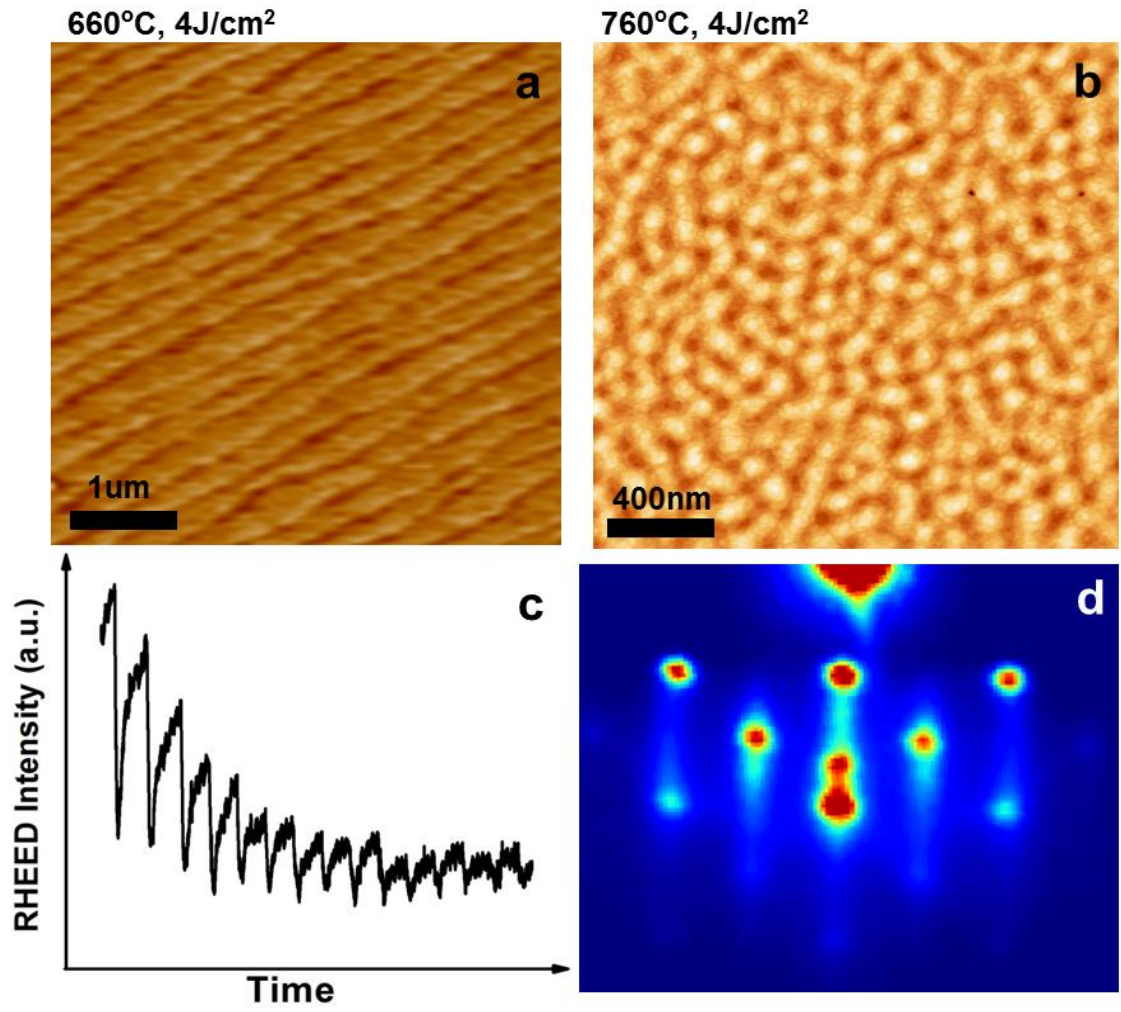

Figure 2.

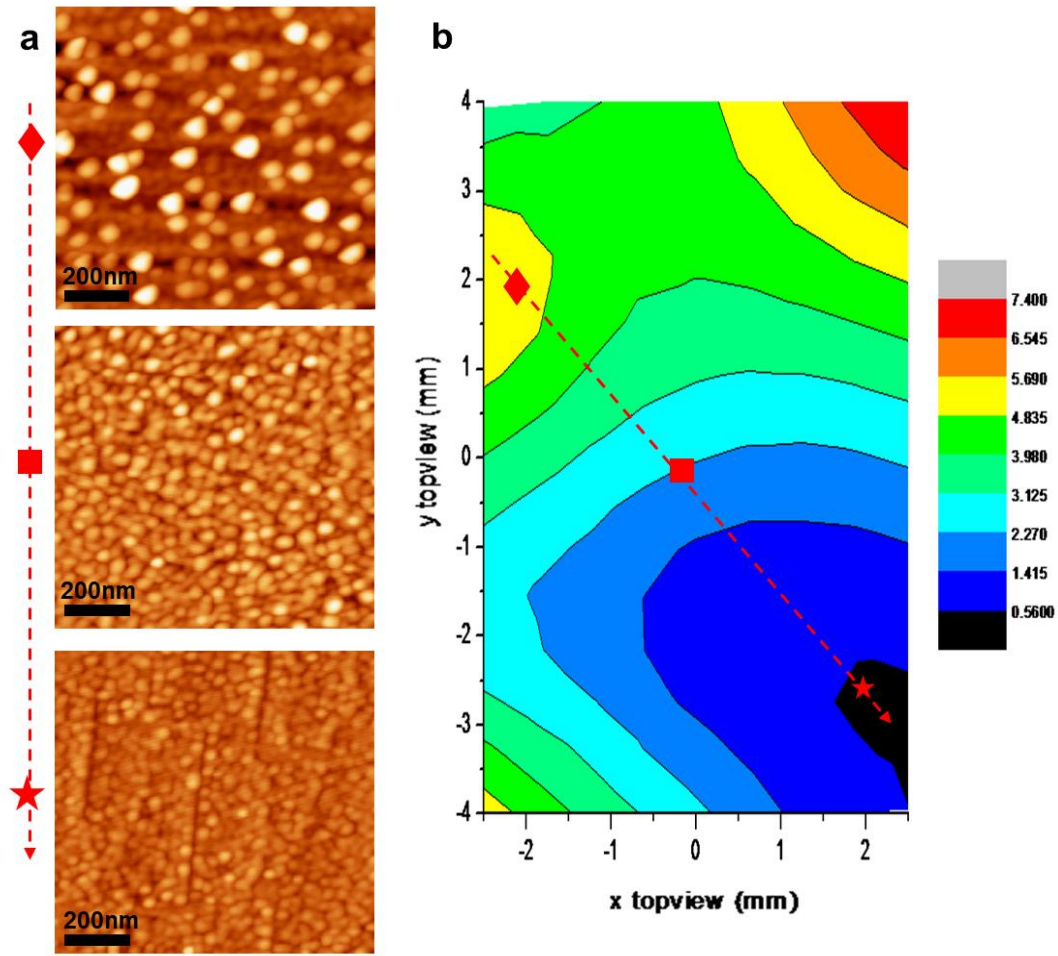

Figure 3.

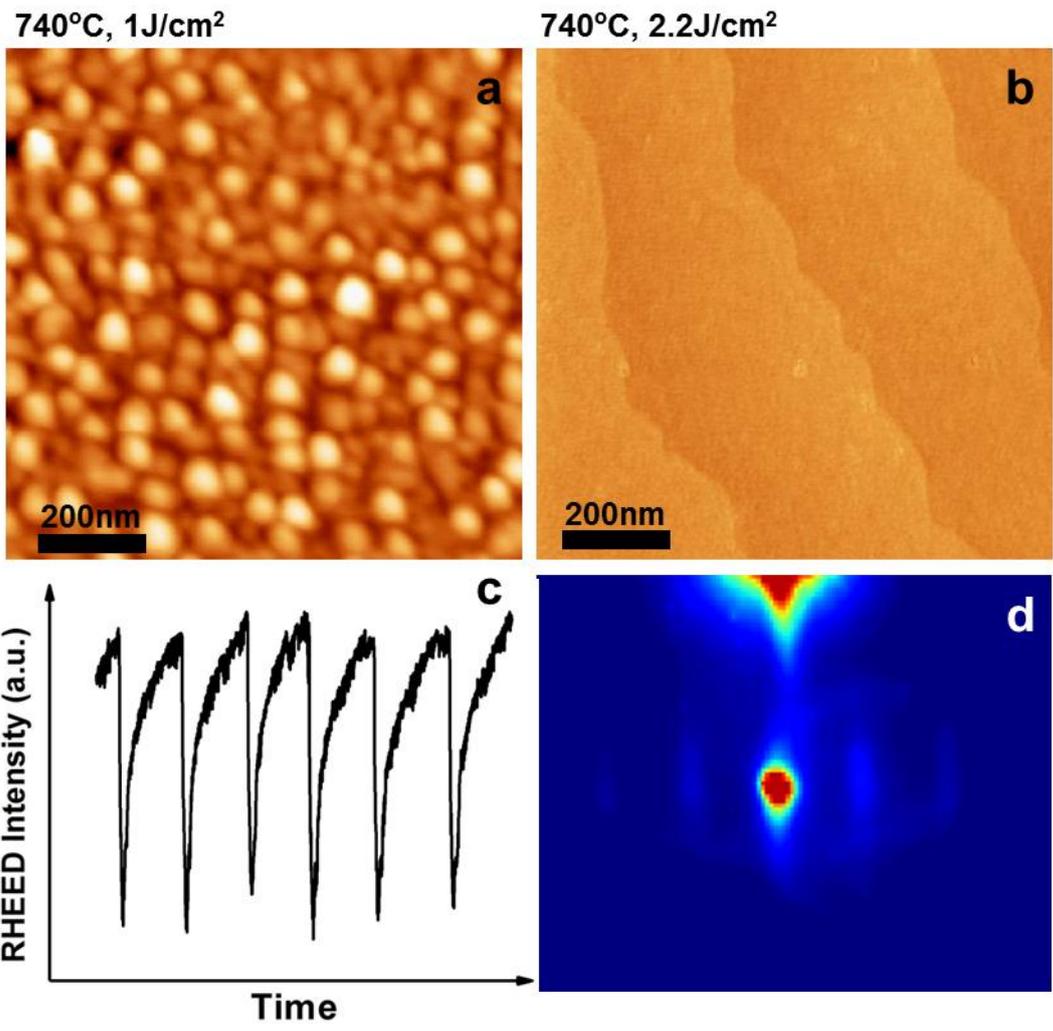

Figure 4.

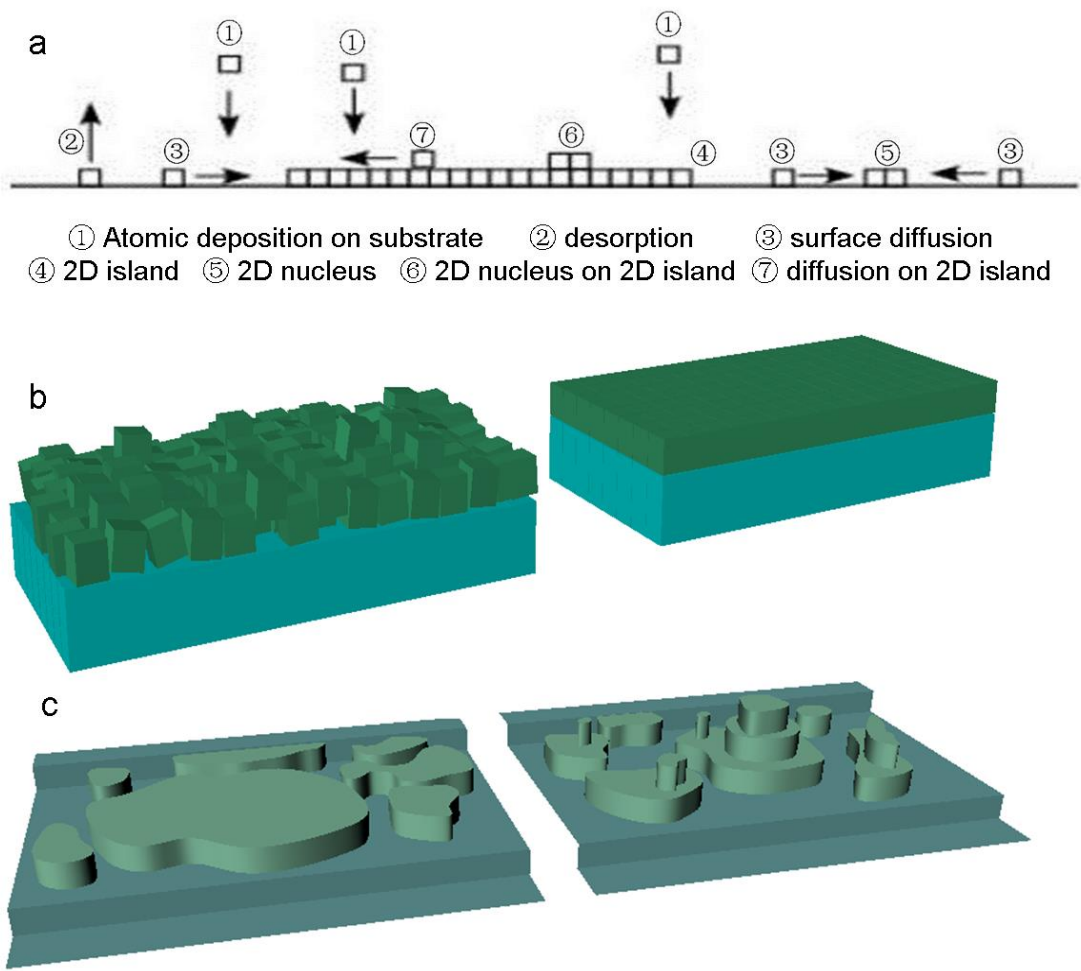

Figure 5.

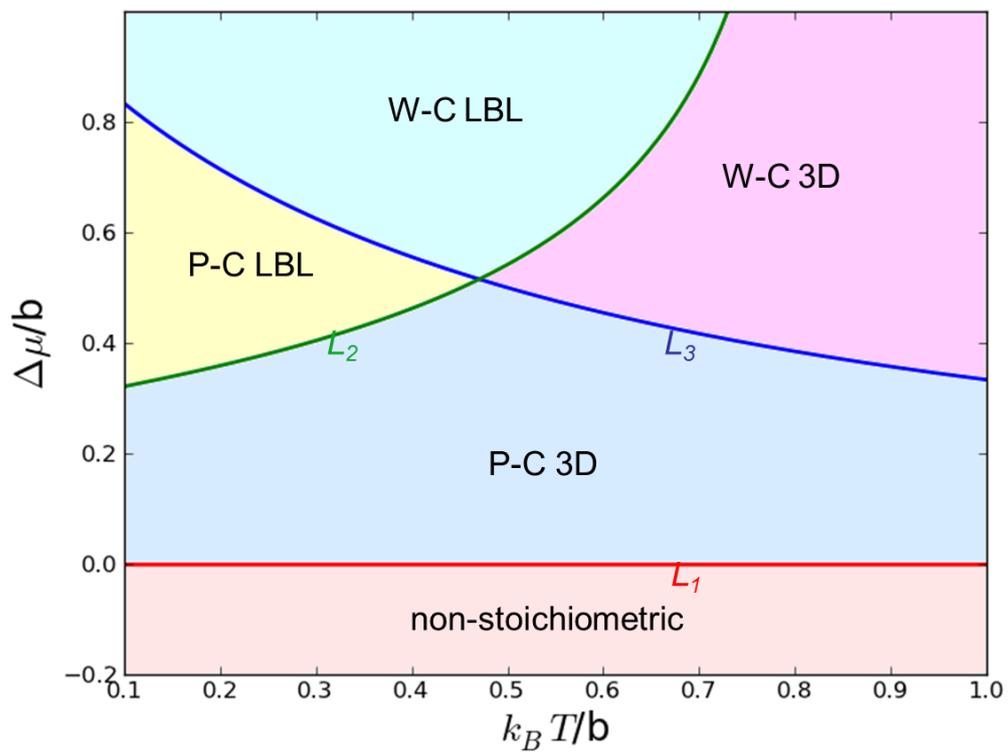

Figure 6.

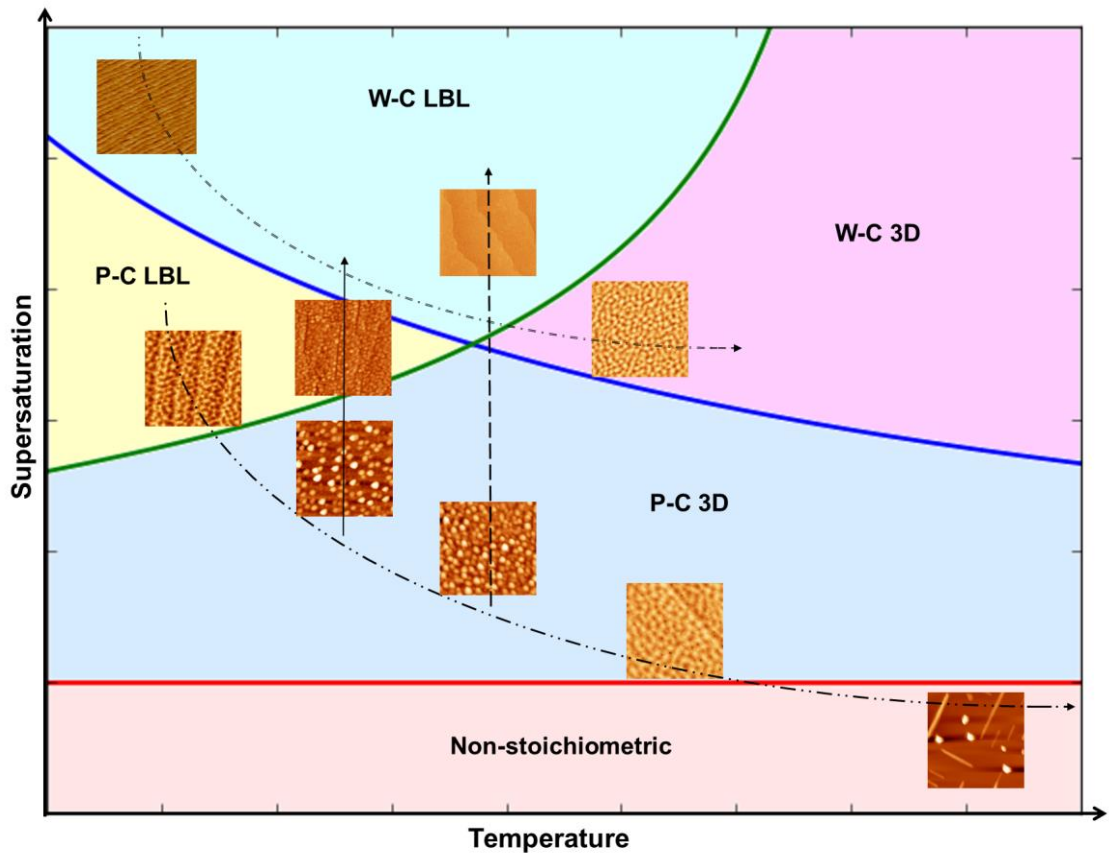

Figure 7.

|  | Counts of film | Counts of film and rod | Counts of rod | Standard Relative Intensity | Counts of rod normalized with standard intensity |
|---|---|---|---|---|---|
| **Mn (Kα)** | 52 | 120 | 68±9 | 150 | 0.45±0.06 |
| **O (K)** | 470 | 529 | 59±22 | 151 | 0.39±0.14 |

**Table 1**. EDX analysis of non-stoichiometric sample (840 $^o$C, 1J/cm$^2$).